\pdfoutput=1
\documentclass[preprint]{aastex701}

\usepackage{CJKutf8}
\usepackage{booktabs}
\usepackage{subcaption}
\usepackage{amsmath}
\usepackage{mathrsfs}
\usepackage{graphicx}
\usepackage{tabularx}
\usepackage{txfonts}
\usepackage{bm}

\received{\today}
\revised{2026}
\accepted{DRAFT}

\submitjournal{ApJL}

\begin{document}
\begin{CJK}{UTF8}{bsmi}
\begin{CJK}{UTF8}{mj}

\title{
Nucleus and Postperihelion Activity of Interstellar Object 3I/ATLAS Observed by Hubble Space Telescope
}

\author[orcid=0000-0001-9067-7477, gname=Man-To, sname=Hui]{Man-To Hui (許文韜)}
\affiliation{
Shanghai Astronomical Observatory, Chinese Academy of Sciences,
No. 80 Rd Nandan, Shanghai 200030, Mainland China
}
\email[show]{mthui@shao.ac.cn, manto@hawaii.edu}

\author[gname=David, sname=Jewitt]{David Jewitt} 
\affiliation{
Department of Earth, Planetary, and Space Sciences, UCLA,
595 Charles Young Drive East, Los Angeles, CA 90095-1567, USA
}
\email{jewitt@ucla.edu}

\author[orcid=0000-0002-0088-3021, gname=Max, sname=Mutchler]{Max J. Mutchler}
\affiliation{
Space Telescope Science Institute, 3700 San Martin Drive, Baltimore, MD 21218, USA
}
\email{mutchler@stsci.edu}

\author[sname=Jessica, gname=Agarwal]{Jessica Agarwal}
\affiliation{
Institut f{\"u}r Geophysik und extraterrestrische Physik, Technische Universit{\"a}t Braunschweig,
Mendelssohnstr. 3, 38106 Braunschweig, Germany
}
\email{j.agarwal@tu-braunschweig.de}

\author[orcid=0000-0002-4676-2196, sname=Yoonyoung, gname=Kim]{Yoonyoung Kim (김윤영)}
\affiliation{
Department of Earth, Planetary, and Space Sciences, UCLA,
595 Charles Young Drive East, Los Angeles, CA 90095-1567, USA
}
\email{yoonyoung@epss.ucla.edu}

\begin{abstract}

We report the detection of the nucleus of interstellar object 3I/ATLAS, using a nucleus extraction technique on Hubble Space Telescope (HST) observations taken between December 2025 and January 2026. The product of the V-band geometric albedo, $p_V$, with the physical cross-section of the nucleus is $0.22 \pm 0.07$ km$^{2}$, which corresponds to an effective radius of $1.3 \pm 0.2$ km if assuming a comet-like albedo $p_{V} = 0.04$. This size is in agreement with an independent estimate based on the reported nongravitational acceleration and activity of the interstellar object. If the measured photometric variations are solely due to the rotation of an aspherical nucleus, the axis ratio must be $2:1$ or greater, and the rotation period $\ga\!1$ hr. Leveraging the range of covered phase angles, we identified a significant opposition surge of $\sim\!0.2$ mag with a width of $3\degr \pm 1\degr$, which may include concurrent contributions from orbital plane crossing and tail projection, and determined a linear phase slope of $0.026 \pm 0.006$ mag degree$^{-1}$ for the coma dust. Compared to the preperihelion brightening trend, 3I faded more rapidly on the outbound leg, following an activity index of $4.5 \pm 0.3$, not unusual in the context of solar system comets. This activity asymmetry is further corroborated by a postperihelion coma surface brightness profile that is significantly shallower than its preperihelion counterpart. From discovery statistics, we infer that multiple interstellar objects resembling 3I probably went undetected prior to the discovery of 1I/`Oumuamua, unless the overall population possesses a steep size distribution.

\end{abstract}


\section{Introduction}
\label{sec:intro}

3I/ATLAS, formerly designated C/2025 N1 (ATLAS), was discovered using the Asteroid Terrestrial-impact Last Alert System (ATLAS) sky survey on 1 July 2025. Serendipitous prediscovery observations were quickly identified \citep[e.g.][] {2025ApJ...991L...2F,2025ApJ...994L..51M,2025ApJ...993L..31Y}, along with numerous followup observations, leading to the definitive identification of the object as the third interstellar object known to intrude into our solar system \citep{2025MPEC....N...12D}. Its trajectory is significantly more hyperbolic with respect to the Sun than those of the previous two interstellar objects, 1I/`Oumuamua and 2I/Borisov, with an eccentricity of $e = 6.14$. The closest approach of the object to the Sun occurred on 29 October 2025 at a perihelion distance of $q = 1.4$ au, implying an excess speed of $\sim\!58$ km s$^{-1}$, which suggests a kinematic age of 3-11 Gyr for 3I, older than the previous interstellar objects \citep{2025ApJ...990L..14T}. The object is statistically unlikely to have approached any star as closely as the current approach to the Sun \citep{2025ApJ...994L...3J}. 

Numerous observations of 3I have been conducted using both ground-based facilities and spacecraft, spanning a wide range of wavelengths and utilising various techniques including imaging, spectroscopy, and polarimetry.
In brief the interstellar object was reported to exhibit a steady growth of activity as it approached  the Sun \citep[e.g.][]{2025arXiv251222354C,2025ApJ...994L...3J,2025A&A...702L...3S}, with peculiar dust scattering properties \citep{2025ApJ...992L..29G, 2026arXiv260108591C}, slightly enriched volatile content, and distinct outgassing behaviour \citep[e.g.][]{2025ApJ...991L..43C,2025arXiv250926053H,2025arXiv251120845R}. Its activity was dominated by CO$_{2}$ at heliocentric distances $r_{\rm H} \ga 3$ au, which was later overtaken by H$_{2}$O sublimation when closer to the Sun \citep[e.g.][]{2025arXiv251222354C,2025ApJ...991L..43C,2025ApJ...991L..50X,2025ApJ...992L...9Y}. However, there has been no successful observation of the nucleus of 3I to date, primarily because of hindrance by its activity, which began prior to the earliest available observations \citep[e.g.][]{2025ApJ...993L..31Y}. Our previous attempt to extract the nucleus signal of 3I from preperihelion Hubble Space Telescope (HST) observations  in July 2025 yielded  an upper limit of $\sim\!2.8$ km for the nucleus radius \citep{2025ApJ...990L...2J}. Nevertheless, characterising the radius of the nucleus remains scientifically pivotal, as an essential input for physical models and  a constraint on the population of interstellar objects as a whole. In this letter, we report the first detection of the nucleus of 3I using postperihelion optical imaging from our HST program.  

\section{Observations}
\label{sec:obs}

\begin{deluxetable*}{lccccccccccc}
\tablecaption{
Observing Geometry of 3I/ATLAS from HST
\label{tab:vgeo}}
\tablewidth{0pt}
\tablehead{
\multicolumn{4}{c}{HST Observations} &
\multicolumn{8}{c}{Observing Geometry}
\\
\cmidrule(lr){1-4}
\cmidrule(lr){5-12}
\colhead{Visit} & 
\colhead{Obs. Time \& Date (UTC)\tablenotemark{a}} &
\colhead{$t_{\rm exp}$ (s)\tablenotemark{b}} & 
\colhead{\#\tablenotemark{c}} & 
\colhead{$r_{\rm H}$ (au)\tablenotemark{d}} & 
\colhead{$\Delta$ (au)\tablenotemark{e}} & 
\colhead{$\alpha$ (\degr)\tablenotemark{f}} & 
\colhead{$\varepsilon$ (\degr)\tablenotemark{g}} & 
\colhead{$\theta_{-\odot}$ (\degr)\tablenotemark{h}} & 
\colhead{$\theta_{-{\bf V}}$ (\degr)\tablenotemark{i}} & 
\colhead{$\psi$ (\degr)\tablenotemark{j}} & 
\colhead{$\nu$ (\degr)\tablenotemark{k}}
}
\startdata
1\tablenotemark{$\dagger$} & 20:39-21:16 30 Nov 2025 & 260 & 6 & 
1.798 & 1.913 & 30.6 & 68.2 & 293.7 & 109.9 & 2.4 & 44.4
\\
2 & 21:22-21:52 12 Dec 2025 & 170 & 6 & 
2.108 & 1.814 & 27.8 & 93.0 & 293.0 & 109.4 & 2.3 & 54.2 
\\
3 & 16:00-16:24 27 Dec 2025 & 170 & 5 & 
2.536 & 1.830 & 18.3 & 126.0 & 291.4 & 107.0 & 1.9 & 62.7 
\\
4 & 15:32-16:08 07 Jan 2026 & 170 & 7 & 
2.874 & 1.978 & 9.8 & 150.3 & 290.6 & 104.1 & 1.4 & 67.3
\\
5 & 15:23-15:54 14 Jan 2026 & 170 & 6 &
3.095 & 2.135 & 4.8 & 164.7 & 293.1 & 102.0 & 1.0 & 69.7
\\
6 & 13:12-13:42 22 Jan 2026 & 170 & 6 &
3.348 & 2.364 & 0.7 & 177.6 & 12.4$\rightarrow$13.5 & 99.6 & 0.7 & 72.0
\enddata
\tablenotetext{a}{Mid-exposure epoch of observation.}
\tablenotetext{b}{Individual exposure time.}
\tablenotetext{c}{Number of individual exposures.}
\tablenotetext{d}{Heliocentric distance.}
\tablenotetext{e}{HST-centric distance.}
\tablenotetext{f}{Phase angle.}
\tablenotetext{g}{Solar elongation.}
\tablenotetext{h}{Position angle of antisolar direction projected in the sky plane of HST.}
\tablenotetext{i}{Position angle of negative heliocentric velocity projected in the sky plane of HST.}
\tablenotetext{j}{Orbital plane angle. Positive values indicate HST 
being above the orbital plane of 3I.}
\tablenotetext{k}{True anomaly of the heliocentric orbit.}
\tablenotetext{\dagger}{Data not included for analysis due to the overexposure of 3I.}
\end{deluxetable*}

Our postperihelion observations of interstellar object 3I were conducted using the 2.4 m HST and the UVIS channel of the Wide Field Camera 3 (WFC3) camera. Data were acquired through the broadband F350LP filter from late 2025 to early 2026 under the Director's Discretionary (DD) program 18152. The filter, characterised by a central wavelength of 5846 \AA~and a full-width-half maximum (FWHM) of 4758 \AA, enabled us to fully leverage the maximal sensitivity of the facility. To reduce overheads we used only the $2047 \times 2050$ full quadrant UVIS2-2K2C-SUB aperture for readout, which provides a nearly square field of view of $\sim\!81\arcsec \times 81\arcsec$ at a pixel scale of 0\farcs04 pixel$^{-1}$. During each HST visit, we collected multiple consecutive exposures tracking the ephemeris position of 3I. The observation configurations and corresponding observing geometries of 3I are summarised in Table \ref{tab:vgeo}.

\section{Nucleus Size Analysis}
\label{sec:an}

The charge transfer efficiency (CTE) of the UVIS channel degrades over time, and therefore, appropriate corrections are necessary to mitigate signal losses (see the \href{https://www.stsci.edu/files/live/sites/www/files/home/hst/instrumentation/wfc3/_documents/wfc3_ihb.pdf}{Wide Field Camera 3 Instrument Handbook}). We started with the calibrated, flat-fielded, and CTE-corrected HST data from our program, retrieved from the Barbara A. Mikulski Archive for Space Telescopes (MAST) \href{https://mast.stsci.edu/portal/Mashup/Clients/Mast/Portal.html}{site}. Cosmic rays were eliminated using the Laplacian cosmic-ray rejection algorithm {\tt L.A.Cosmic} \citep{2001PASP..113.1420V} for the Image Reduction and Analysis Facility \citep[{\tt IRAF};][]{1986SPIE..627..733T}.\footnote{The code is publicly available at the \href{http://www.astro.yale.edu/dokkum/lacosmic/}{L.A.Cosmic website}.} Regions free of cosmic-ray contamination remained untouched by this process, with the only exception that overexposed star trails were mistakenly modified by {\tt L.A.Cosmic}. This had no impact on our result, however, as none of these trails were in the vicinity of the photocenter of the interstellar object.

Prior to performing photometry, we utilised the \href{https://etc.stsci.edu/etc/input/wfc3uvis/imaging/}{WFC3 UVIS Imaging Exposure Time Calculator} and the stellar spectral flux library by \citet{1998PASP..110..863P} covering possible spectral slopes of long-period comets in the solar system \citep[e.g.][]{2015AJ....150..201J} to simulate the image zero-point for objects of similar colours. The resulting scatter was treated as the uncertainty associated with the image zero-point. 



\subsection{Conservative Constraint}
\label{ssec:cc}

The bright and centrally condensed coma of 3I precluded direct measurement of the nucleus. We therefore attempted to isolate and remove the coma contribution to the photometry of the nucleus in different ways. The simplest method is to treat the coma adjacent to the photocentric region as the foreground (background in photometry), which inevitably underestimates the contribution of the coma in the near-nucleus region, thereby yielding an upper limit for the nucleus signal. 
We measured the total flux within a circular aperture 0\farcs16 (4 pixels) in radius at the photocenter of 3I, from which we subtracted  the background flux within a contiguous annulus  extending to 0\farcs24 (6 pixels) from the photocenter. These flux measurements were then converted to apparent magnitudes using the simulated image zero-point. 

The apparent magnitudes are functions of heliocentric distance ($r_{\rm H}$), HST-centric distances ($\Delta$), and phase angle ($\alpha$), all of which evolved over the course of our HST observations. We normalised both distances to $r_{\oplus} = 1$ au and corrected for phase effects by assuming a linear phase model with a slope of $\beta_{\alpha} = 0.04 \pm 0.02$ mag deg$^{-1}$ \citep[e.g.][]{2004come.book..223L,2024come.book..361K} to compute the absolute $V$-band magnitude of the nucleus:
\begin{equation}
H_{{\rm n}, V} = m_{{\rm n}, V} - 5 \log \left( r_{\rm H} \Delta \right) - \beta_{\alpha} \alpha
\label{eq:Hnuc},
\end{equation}
\noindent where $m_{{\rm n}, V}$ is the apparent $V$-band magnitude of the nucleus, $r_{\rm H}$ and $\Delta$ are expressed in au, and $\alpha$ is phase angle in degrees. We thereby determined a lower bound of $H_{{\rm n}, V} \ga 15$ across the five HST epochs for the nucleus of the interstellar object, consistent with $H_{{\rm n}, V} \ga 15.4$ from the preperihelion HST measurement \citep{2025ApJ...990L...2J}.

Reformulated from \citet{1991ASSL..167...19J}, the optical cross-section of the nucleus at zero phase angle, denoted $C_{\rm n}$, is related to the absolute magnitude by the following equation:
\begin{align}
\nonumber
C_{\rm n} & \triangleq p_{V} \Xi_{\rm n}
\\
& = 10^{0.4 \left(m_{\odot, V} - H_{{\rm n}, V} \right)} \pi r_{\oplus}^{2}
\label{eq:pR2},
\end{align}
\noindent where $p_{V}$ is the $V$-band geometric albedo, $\Xi_{\rm n}$ is the effective geometric cross-section ($\Xi_{\rm n} = \pi R_{\rm n}^{2}$ for a spherical nucleus of radius $R_{\rm n}$) of the nucleus, and $m_{\odot, V} = -26.76 \pm 0.03$ is the apparent $V$-band magnitude of the Sun as observed at a heliocentric distance of $r_{\oplus} = 1$ au \citep{2018ApJS..236...47W}. Substitution yields $C_{\rm n} \la 2$ km$^2$ for all HST epochs. Assuming a spherical shape and a nominal value of $p_{V} = 0.04$ typical for cometary nuclei \citep[e.g.][]{2004come.book..223L,2024come.book..361K}, but as yet unconstrained for 3I, we derived a conservative upper limit estimate for the nucleus radius of 3I of $R_{\rm n} \la 4$ km.

\subsection{Nucleus Extraction}
\label{ssec:nuc_ext}

We next applied a nucleus extraction technique originally devised by \citet{1995A&A...293L..43L} and implemented by \citet{1998A&A...337..945L}. By employing this technique, we successfully detected the nucleus of long-period comet C/2014 UN$_{271}$ (Bernardinelli-Bernstein) from our HST program using similar observing configurations \citep{2022ApJ...929L..12H}. Basically, the technique assumes an optically thin coma and  decomposes the total surface brightness into distinct contributions from the nucleus and coma as
\begin{align}
\nonumber
\Sigma \left(\bm{\rho} \right) & = \left[k_{\rm n} \delta\left(\bm{\rho} \right) + k_{\rm c} \left( \theta \right) \left(\dfrac{\rho}{\rho_{0}} \right)^{-\gamma \left( \theta \right)} \right] \ast \mathcal{P}\left( \bm{\rho} \right)
\\
& = k_{\rm n} \mathcal{P} \left(\bm{\rho} \right) + \left[k_{\rm c} \left( \theta \right) \left(\dfrac{\rho}{\rho_{0}} \right)^{-\gamma \left( \theta \right)} \right] \ast \mathcal{P} \left(\bm{\rho} \right)
\label{eq:sprof},
\end{align}
\noindent where the observed total surface brightness, $\Sigma$, is a function of the projected vector from the photocenter, $\bm{\rho}$, having length $\rho$ and azimuthal angle $\theta$. The intrinsic nucleus signal is modelled as a two-dimensional Dirac delta function $\delta$, scaled by a constant factor $k_{\rm n}$. The coma contribution is parametrised by scaling factor $k_{\rm c}$ and logarithmic surface brightness gradient $\gamma$, both as functions of the azimuthal angle. A spherically symmetric coma in steady state has $\gamma = 1$, whereas solar radiation pressure steepens the gradient to $\gamma = 1.5$ in the antisunward direction \citep{1987ApJ...317..992J}. The observed signal is the result of convolving the intrinsic profile with the WFC3 camera PSF ($\mathcal{P}$), which we simulated using the {\tt TinyTim} package \citep{2011SPIE.8127E..0JK}. In Equation \eqref{eq:sprof} the symbol $\ast$ denotes the convolution operator. We adopted a normalisation distance of $\rho_{0} =1$ pixel to ensure that the scaling factors for the nucleus and the coma share the same units.

\begin{figure*}
\centering
\begin{subfigure}[b]{.495\textwidth}
\centering
\includegraphics[width=\textwidth]{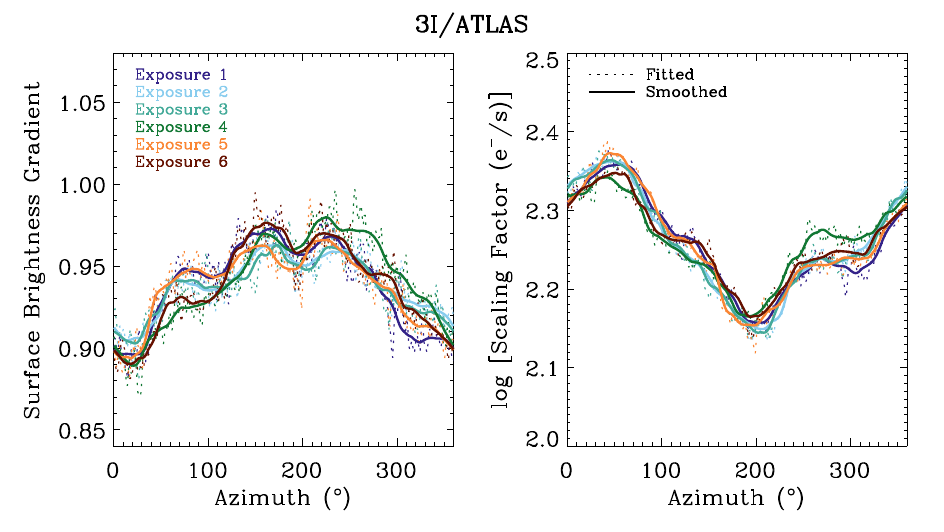}
\caption{
Visit 2
\label{fig:coma_pars_v2}
}
\end{subfigure}
\begin{subfigure}[b]{.495\textwidth}
\centering
\includegraphics[width=\textwidth]{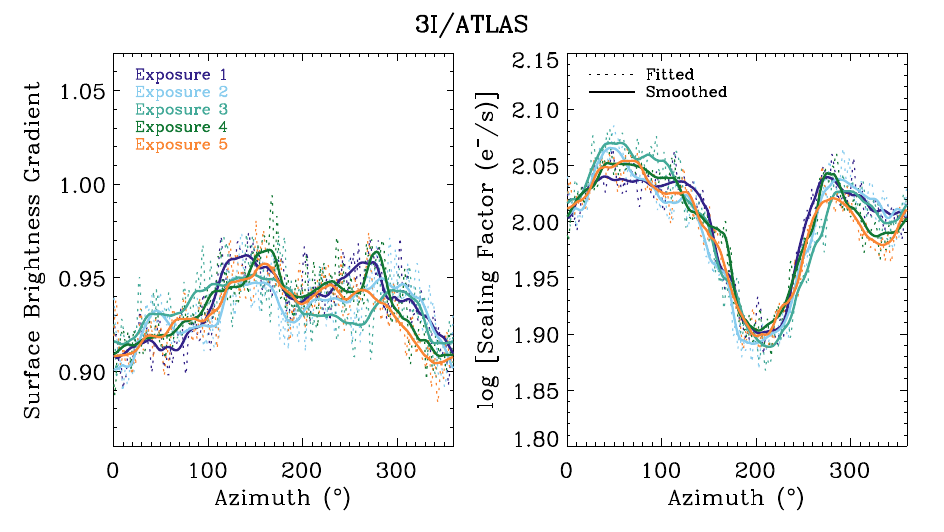}
\caption{
Visit 3
\label{fig:coma_pars_v3}
}
\end{subfigure}
\begin{subfigure}[b]{.495\textwidth}
\centering
\includegraphics[width=\textwidth]{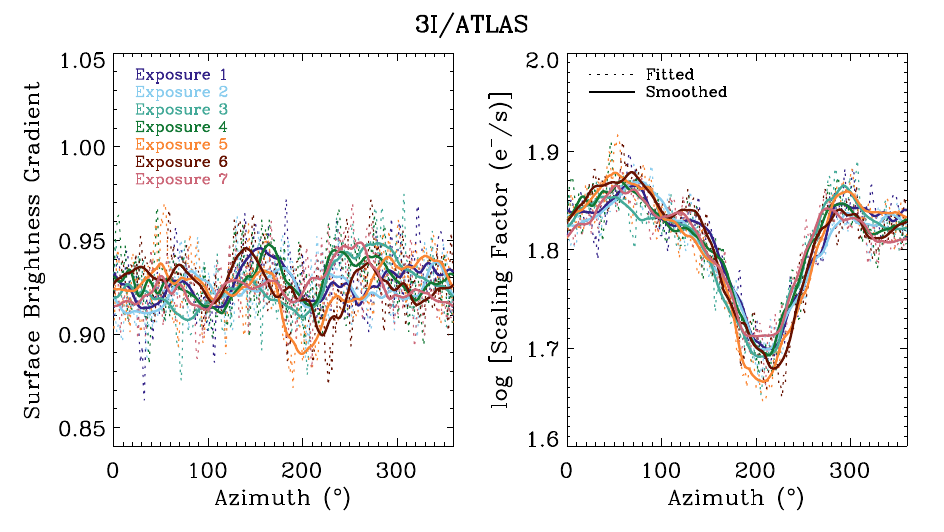}
\caption{
Visit 4
\label{fig:coma_pars_v4}
}
\end{subfigure}
\begin{subfigure}[b]{.495\textwidth}
\centering
\includegraphics[width=\textwidth]{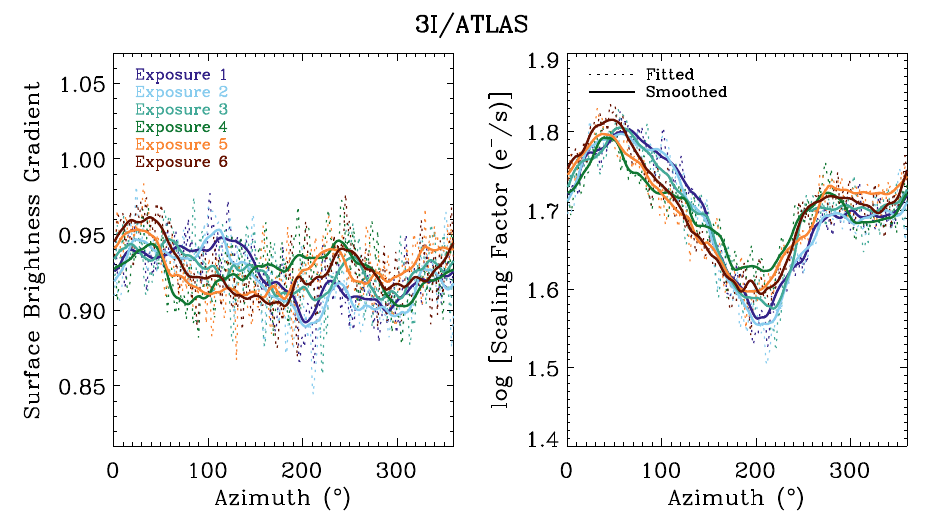}
\caption{
Visit 5
\label{fig:coma_pars_v5}
}
\end{subfigure}
\begin{subfigure}[b]{.495\textwidth}
\centering
\includegraphics[width=\textwidth]{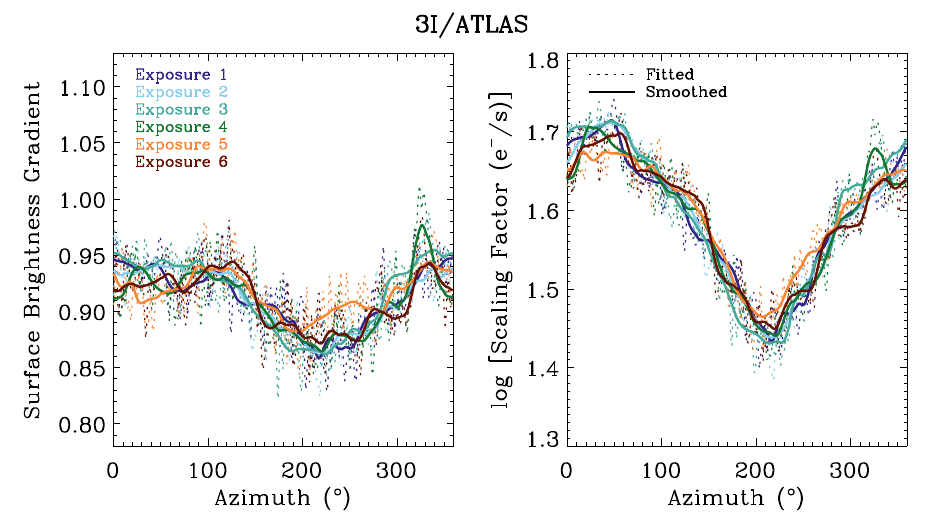}
\caption{
Visit 6
\label{fig:coma_pars_v6}
}
\end{subfigure}
\caption{
Best-fit parameters for the surface brightness profile of the coma as functions of azimuth (measured from J2000 celestial equatorial north in an anticlockwise direction). The fits were performed within an annular region between 6 pixels (0\farcs24) and 30 pixels (1\farcs20) from the photocenter of 3I/ATLAS. Panels represent results from (a) Visit 2 on 12 December 2025, (b) Visit 3 on 27 December 2025, (c) Visit 4 on 7 January 2026, (d) Visit 5 on 14 January 2026, and (e) Visit 6 on 21 January 2026. Data from individual exposures are color coded. Solid lines represent the results of a median smoothing applied to the best fits with an azimuthal window of 30\degr.
\label{fig:coma_pars}
}
\end{figure*}

\begin{figure*}
\epsscale{1.0}
\plotone{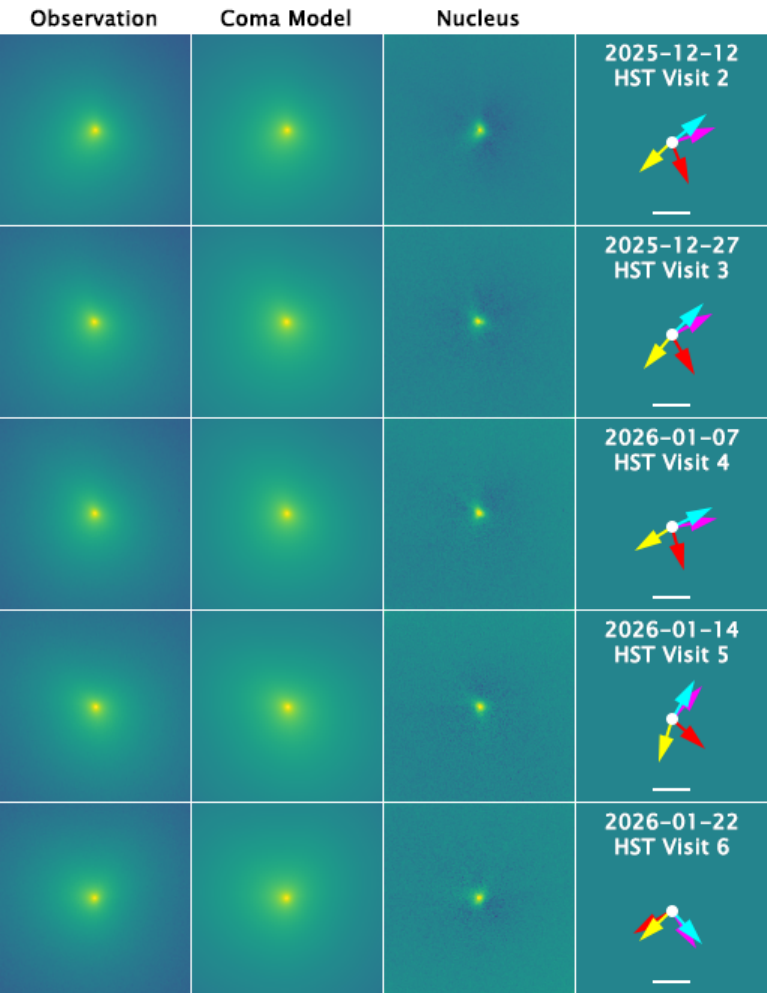}
\caption{
Extraction of the nucleus of 3I/ATLAS. The observed HST image (first panel from the left) is shown alongside the best-fit coma model (second panel) for each HST visit (indicated at right). Subtraction of the model from the observation reveals the nucleus of 3I (third panel). For illustrative purposes, the displayed images are median-combined from individual exposures within each visit and displayed with a logarithmic intensity stretch. In each row, the red and magenta arrows indicate local north and east, respectively, in the J2000 equatorial coordinate system, with the projected antisolar direction and the negative heliocentric velocity of 3I represented by the yellow and cyan arrows, respectively. The horizontal white bar near the bottom marks a scale of 1\arcsec~in apparent length.
\label{fig:obs_mdl_nuc}}
\end{figure*}

Following \citet{2018PASP..130j4501H} we assigned a weighting map based on Poisson statistics to each individual exposure and utilised the Levenberg-Marquardt algorithm via {\tt MPFIT} \citep{2009ASPC..411..251M} to determine the best-fit surface brightness profiles for the coma. The fits were restricted to an annular region between 0\farcs24 and 1\farcs20 in radius from the photocenter of 3I. The results are shown as functions of azimuth in Figure \ref{fig:coma_pars}, where sharp fluctuations are caused by residual cosmic-ray artifacts. We performed azimuthal median smoothing on these fits (shown as solid curves in Figure \ref{fig:coma_pars}) and applied the resulting parameters to reconstruct the surface brightness profile of the coma on a sevenfold-subsampled grid. This reconstruction was extended into the near-nucleus region, which is interior to the fitted annulus, assuming that the coma brightness therein follows the same profile trend as within the annulus. As shown in the Figure, the surface brightness profile index was slightly $<$1, contrasting with values $>$1 in the preperihelion HST data \citep{2025ApJ...990L...2J}. After subtracting the coma model from each exposure and resampling the resulting image back to the original pixel scale, a distinct positive residual feature was revealed at the original photocentric position. The FWHM of this feature is statistically consistent with the WFC3 PSF model generated by {\tt TinyTim}, although faint jet-like features remain. Consequently, we interpret this residual as the detected nucleus of 3I. Figure \ref{fig:obs_mdl_nuc} displays the observed, modelled, and residual images of the interstellar object for each HST visit.

\begin{figure*}
\centering
\begin{subfigure}[b]{.33\textwidth}
\centering
\includegraphics[width=\textwidth]{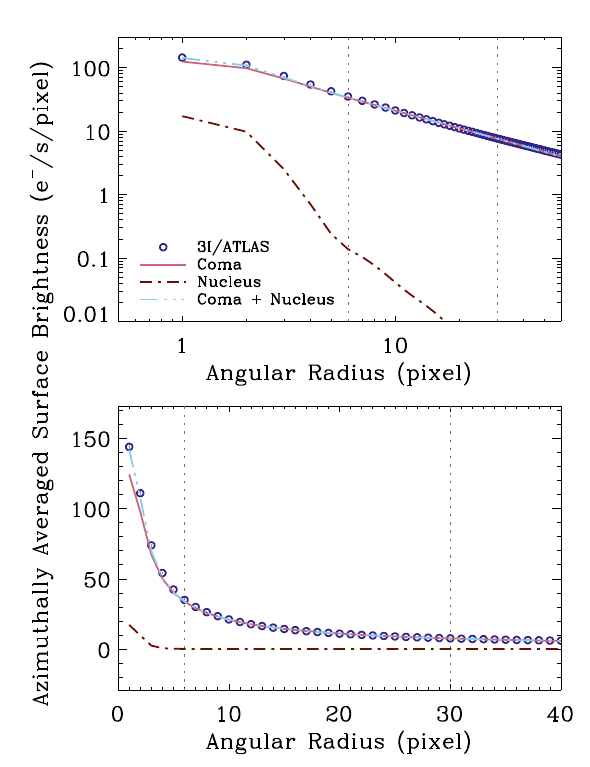}
\caption{
Visit 2
\label{fig:radprof_v2}
}
\end{subfigure}
\begin{subfigure}[b]{.33\textwidth}
\centering
\includegraphics[width=\textwidth]{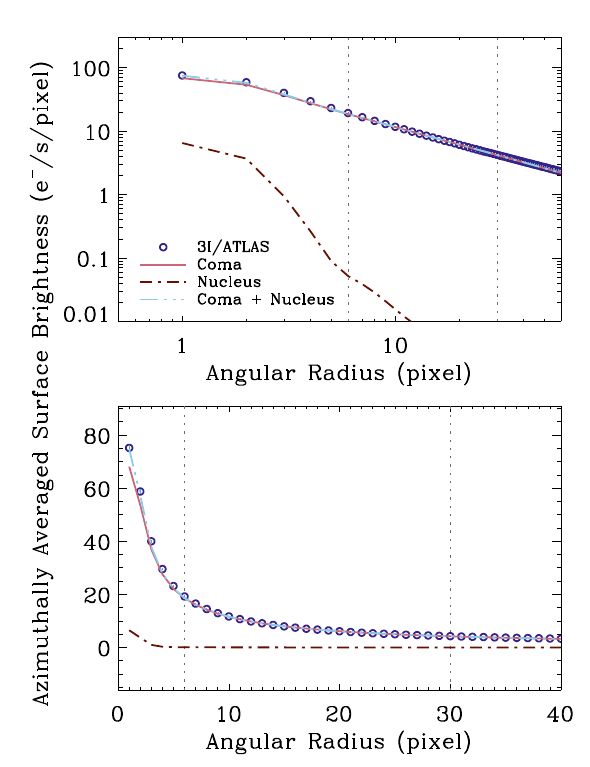}
\caption{
Visit 3
\label{fig:radprof_v3}
}
\end{subfigure}
\begin{subfigure}[b]{.33\textwidth}
\centering
\includegraphics[width=\textwidth]{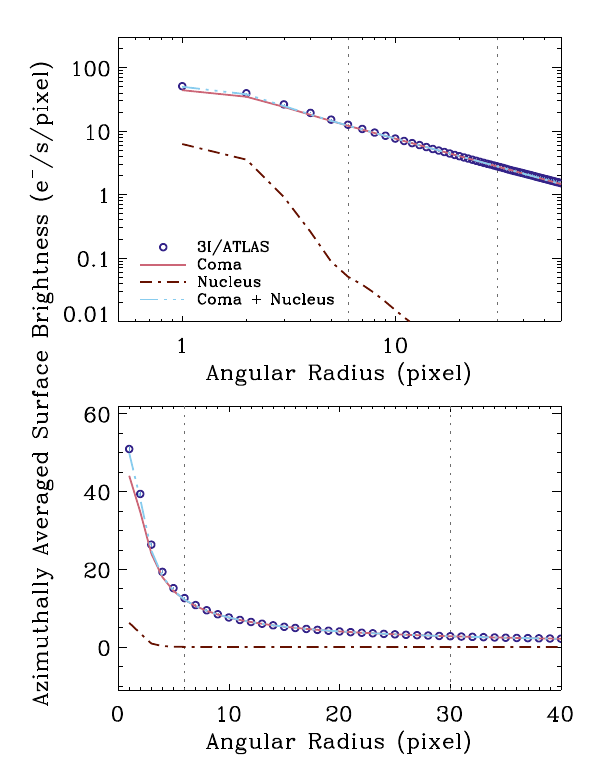}
\caption{
Visit 4
\label{fig:radprof_v4}
}
\end{subfigure}
\begin{subfigure}[b]{.33\textwidth}
\centering
\includegraphics[width=\textwidth]{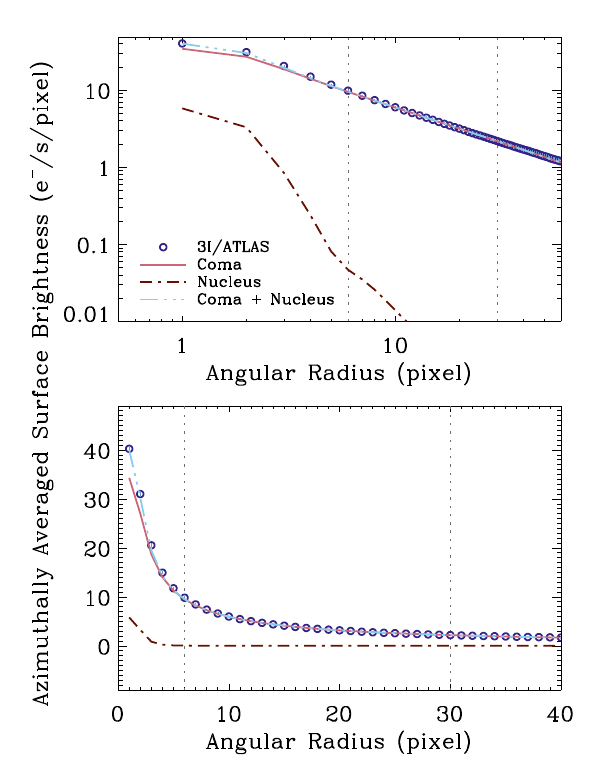}
\caption{
Visit 5
\label{fig:radprof_v5}
}
\end{subfigure}
\begin{subfigure}[b]{.33\textwidth}
\centering
\includegraphics[width=\textwidth]{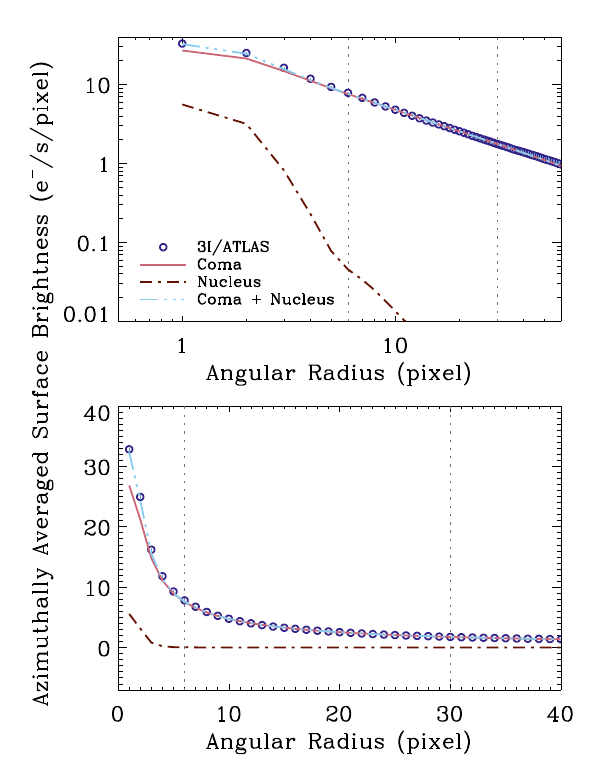}
\caption{
Visit 6
\label{fig:radprof_v6}
}
\end{subfigure}
\caption{
Comparison between the observed radial surface brightness profile (dark red diamonds) and the best-fit model (light blue dashed-dotted line) for the first exposure of 3I/ATLAS from (a) Visits 2 on 12 December 2025, (b) Visit 3 on 27 December 2025, (c) Visit 4 on 7 January 2026, (d) Visit 5 on 14 January 2026, and (e) Visit 6 on 21 January 2026. The best-fit coma and nucleus components are shown as pink solid and dark red dashed-dotted curves, respectively. Formal $1\sigma$ uncertainties are plotted as well but are largely obscured by the markers at the displayed scales. In each panel, the top and bottom sub-panels display the same data on log-log and linear scales, respectively. Grey vertical dotted lines mark the inner and outer boundaries (6 and 30 pixels, or 0\farcs24 and 1\farcs20, respectively) of the annular region used for fitting the coma.  Results for the subsequent exposures are visually consistent and are therefore omitted for clarity. 
\label{fig:radprof}
}
\end{figure*}

\begin{figure*}
\centering
\begin{subfigure}[b]{.33\textwidth}
\centering
\includegraphics[width=\textwidth]{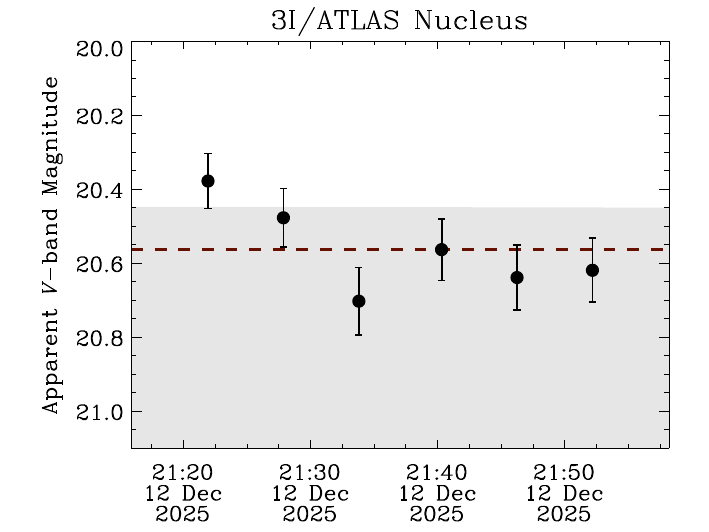}
\caption{
Visit 2
\label{fig:lcn_v2}
}
\end{subfigure}
\begin{subfigure}[b]{.33\textwidth}
\centering
\includegraphics[width=\textwidth]{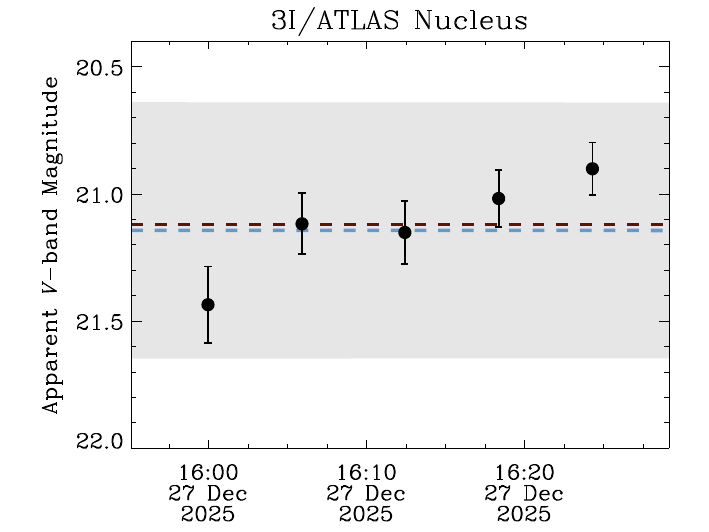}
\caption{
Visit 3
\label{fig:lcn_v3}
}
\end{subfigure}
\begin{subfigure}[b]{.33\textwidth}
\centering
\includegraphics[width=\textwidth]{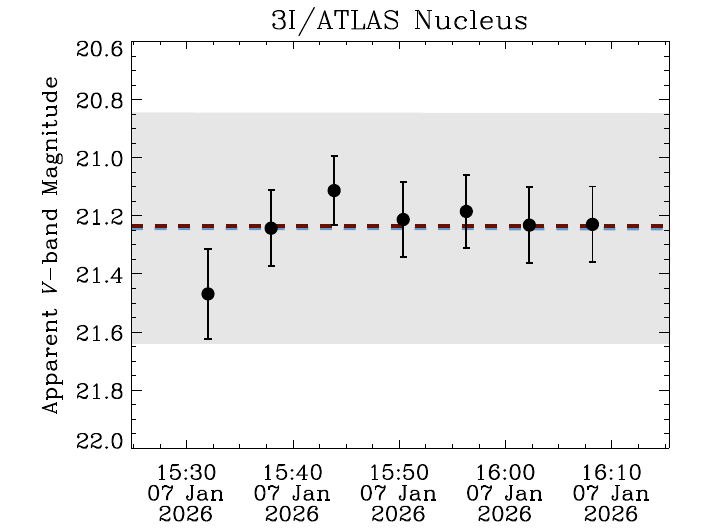}
\caption{
Visit 4
\label{fig:lcn_v4}
}
\end{subfigure}
\begin{subfigure}[b]{.33\textwidth}
\centering
\includegraphics[width=\textwidth]{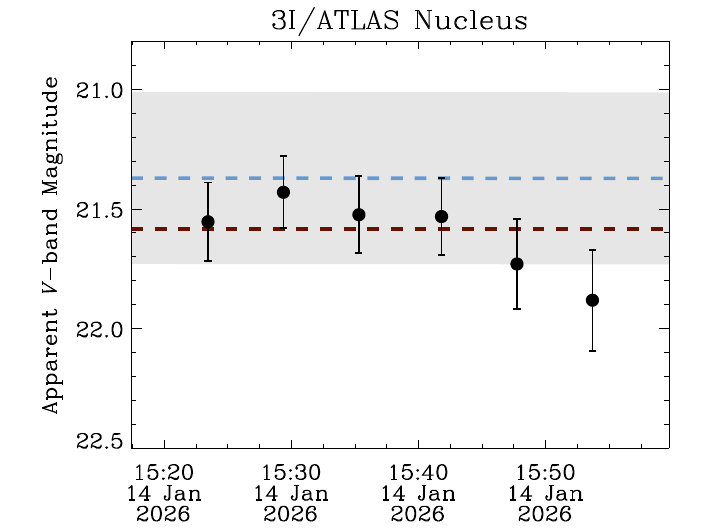}
\caption{
Visit 5
\label{fig:lcn_v5}
}
\end{subfigure}
\begin{subfigure}[b]{.33\textwidth}
\centering
\includegraphics[width=\textwidth]{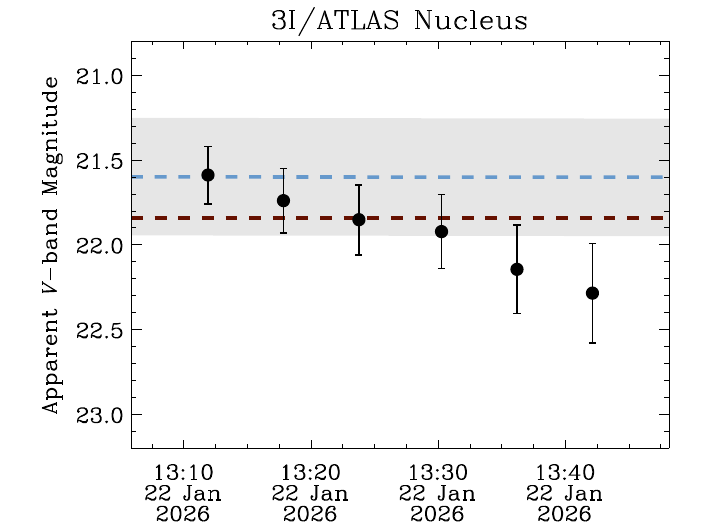}
\caption{
Visit 6
\label{fig:lcn_v6}
}
\end{subfigure}
\caption{
Temporal apparent magnitude variations of the nucleus of 3I across the HST visits. In each panel, the horizontal red dashed line marks the mean value computed from the specific visit. The blue dashed line represents the predicted apparent magnitude of a spherical nucleus of the same effective radius, with its corresponding $1\sigma$ uncertainty region shaded in grey. 
\label{fig:lc_nuc}
}
\end{figure*}

We next performed PSF photometry on the residual signal with {\tt psffit} from \href{https://www2.boulder.swri.edu/~buie/idl/}{Buie's IDL library}.  Results for each HST visit are listed in Table \ref{tab:phot}, while comparisons of the observed and modelled radial brightness profiles and the apparent magnitude time series are shown in Figures \ref{fig:radprof} and \ref{fig:lc_nuc}, respectively. We were fully aware that the extracted nucleus flux possibly fell inside the caution zone where results can be biased by the nucleus extraction technique \citep{2018PASP..130j4501H}. To investigate the reliability of the results, we varied the azimuthal smoothing window width, image subsampling factor, and the fitted annular region within the ranges suggested by \citet{2018PASP..130j4501H}, and we even switched to aperture photometry on the residual feature using an aperture radius corresponding to its FWHM so as to better exclude jet-like features in the near-nucleus region. We found the results to be consistent within the $1\sigma$ significance level regardless of the parameter choices or photometry method. Furthermore, applying the method described in Section \ref{ssec:cc} to the whole image sequence yielded the same temporal trends as those shown in Figure \ref{fig:lc_nuc}, albeit with a systematic brightness offset due to contamination from the coma. Therefore, we conclude that the nucleus extraction technique successfully revealed the nucleus of 3I. Our previous attempt using preperihelion HST observations \citep{2025ApJ...990L...2J} was likely unsuccessful both as a result of lower spatial resolution resulting from a greater (3.0 au) geocentric distance and a steeper preperihelion radial brightness profile of the coma.

Figure \ref{fig:lc_nuc} shows marginal evidence for brightness variability, consistent with the rotation of an aspherical nucleus and/or with residual near-nucleus coma imperfectly removed by the extraction technique. Unfortunately, our data are too sparsely sampled to permit any useful estimate of the nucleus rotation period, except that it is likely $\gtrsim\!1$ hour. Under the assumption that the brightness variations are real and driven purely by rotation, the maximum range of the apparent magnitude within a single HST epoch, $\Delta m_{{\rm n}, V} \ga 0.8$ mag, implies a nucleus axis ratio projected into the plane of the sky of $10^{0.4 \Delta m_{{\rm n}, V}} \ga 2:1$. However, this axis ratio would be lower, even potentially approaching unity, if a significant portion of the variability stems from near-nucleus dust activity.


Averaging these measurements yields an absolute $V$-band magnitude of $H_{{\rm n}, V} = 17.1 \pm 0.4$ for the nucleus of 3I, corresponding to an effective optical cross-section of $C_{\rm n} = 0.22 \pm 0.07$ km$^{2}$. Assuming a $V$-band geometric albedo of $p_{V} = 0.04$, the effective nucleus radius is $R_{\rm n} = 1.3 \pm 0.2$ km, much larger than the nuclei of 1I/`Oumuamua (highly aspherical, with an effective radius of $\sim\!0.08$ km) and 2I/Borisov ($\sim\!0.4$ km) \citep[and citations therein]{2023ARA&A..61..197J}. In Figure \ref{fig:lc_nuc} we denote the predicted apparent magnitude of a spherical nucleus having the same effective mean radius with a blue dashed line to compare it with the time sequence of individual apparent magnitude measurements. The shaded area represents the $1\sigma$ uncertainty region propagated from the standard deviation of the mean and the assumed error in the adopted linear phase function.

\begin{deluxetable*}{lccccccc}
\tablecaption{
Photometry and Physical Properties of 3I/ATLAS Nucleus
\label{tab:phot}}
\tablewidth{0pt}
\tablehead{
\colhead{Visit} &
\colhead{Apparent Magnitude} &
\colhead{Absolute Magnitude} &
\colhead{Optical Cross-section} &
\colhead{Radius\tablenotemark{c}}
\\
& 
$m_{V}$
 & 
$H_{V}$
 &
$C_{\rm n}$ (km$^{2}$)
 &
$R_{\rm n}$ (km)
}
\startdata
2 &
$20.56 \pm 0.03~\left(0.12\right)$ & $16.54 \pm 0.23~\left(0.12 \right)$ &
$0.33 \pm 0.07~\left(0.04 \right)$ & $1.63 \pm 0.17~\left(0.09 \right)$\\
3 & 
$21.12 \pm 0.05~\left(0.18 \right)$ & $17.06 \pm 0.17~\left(0.20 \right)$ &
$0.20 \pm 0.03~\left(0.04 \right)$ & $1.28 \pm 0.10~\left(0.12 \right)$\\
4 &
$21.24 \pm 0.05~\left(0.10 \right)$ & $17.07 \pm 0.09~\left(0.11 \right)$ &
$0.20 \pm 0.02~\left(0.02 \right)$ & $1.28 \pm 0.05~\left(0.06 \right)$\\
5 &
$21.58 \pm 0.07~\left(0.15 \right)$ & $17.29 \pm 0.08~\left(0.15 \right)$ &
$0.16 \pm 0.01~\left(0.02 \right)$ & $1.15 \pm 0.04~\left(0.08 \right)$\\
6 &
$21.84 \pm 0.09~\left(0.24 \right)$ & $17.32 \pm 0.09~\left(0.24 \right)$ &
$0.15 \pm 0.01~\left(0.04 \right)$ & $1.12 \pm 0.05~\left(0.13 \right)$ \\
\hline
\multicolumn{5}{l}{\bf Summary}
\\
\multicolumn{2}{l}{mean (median) $\pm$ standard deviation}
& $17.08~\left(17.07 \right) \pm 0.35$ & $0.22~\left(0.21 \right) \pm 0.07$ & $1.29~\left(1.28 \right) \pm 0.21$
\enddata
\tablecomments{All magnitudes are reported in the $V$ band. We assumed a nominal geometric albedo of $p_{V} = 0.04$ to derive the nucleus radii. For each visit, we provide both the standard error of the weighted mean and the standard deviations, with the latter in parentheses. For the summary line, the format is stated explicitly therein. We applied equal weighting so as to better prevent the mean from being skewed by measurements taken at smaller phase angles. }
\end{deluxetable*}

\begin{deluxetable}{rcl}
\tablecaption{
Comparisons of Nucleus Size Estimates of 3I/ATLAS
\label{tab:sz_nuc_comp}}
\tablewidth{0pt}
\tablehead{
\colhead{Nucleus Radius (km)} &
\colhead{Source} &
\colhead{Method}
}
\startdata
0.26-0.37 & \citet{2025RNAAS...9..329E} & Nongravitational effect \\
0.41-0.53 & \citet{2025arXiv251218341F} & Nongravitational effect \\
$1.3 \pm 0.2$ & \textbf{This work} & Nucleus extraction \\
$1.5 \pm 0.1$\tablenotemark{$\dagger$} & \textbf{This work} & Nongravitational effect \\
$< 2.8$ & \citet{2025ApJ...990L...2J} & Nucleus extraction \\
$> 2.5$ & \citet{2025arXiv250921408C} & Nongravitational effect \\
$\la 4$ & \textbf{This work} & Aperture photometry \\
$\le 6.3 \pm 0.8$\tablenotemark{$\ddagger$} & \citet{2025arXiv250713409C} & Surface profile comparison \\
$< 11 \pm 1$\tablenotemark{$\ddagger$} & \citet{2025ApJ...989L..36S} & Aperture photometry \\
\enddata
\tablenotetext{\dagger}{The collimation coefficient of activity was assumed to be $\kappa = 0.5$. For alternative values of $\kappa \in \left(0, 1 \right)$, the size estimate scales as $\propto \kappa^{1/3}$ per Equation \eqref{eq:Rnuc_NG}. Note that the reported formal error does not take into account the systematic uncertainty in $\kappa$, and hence it is only an underestimate of the actual uncertainty.}
\tablenotetext{\ddagger}{To ensure consistency, estimates originally derived by other authors assuming $p_{V} = 0.05$ have been rescaled to match our adopted value of $p_{V} = 0.04$. }
\end{deluxetable}


\section{Discussion}
\label{sec:disc}

\subsection{Nucleus Size Inferred from Nongravitational Effect}
\label{ssec:sz_nuc_NG}

The nongravitational acceleration of 3I enables an independent estimate of its nucleus size, which can then be compared with the results derived from our HST observations. The \href{https://ssd.jpl.nasa.gov/tools/sbdb_lookup.html#/?sstr=3i}{orbital solution} of 3I by JPL Horizons demonstrates a statistically significant detection of nongravitational accelerations, presumably due to anisotropic mass loss. Adopting an inverse-square law for the nongravitational force model, i.e. $g_{\rm NG} \propto r_{\rm H}^{-2}$, the solution yields the radial, transverse, and normal (RTN) components of the nongravitational parameters, originally defined by \citet{1973AJ.....78..211M}, to be $A_{1} = \left(+4.5 \pm 0.1 \right) \times 10^{-8}$ au day$^{-2}$, $A_{2} = \left(+1.7 \pm 0.1 \right) \times 10^{-8}$ au day$^{-2}$, and $A_{3} = \left(-6.0 \pm 0.2 \right) \times 10^{-9}$ au day$^{-2}$, respectively. These components all exhibit a significance of $\ga\!10 \sigma$.\footnote{Retrieved on 5 February 2026.} Based on the conservation of momentum and assuming a spherical nucleus, the nucleus radius can be inferred from the nongravitational effect and the outgassing mass-loss rate using the formulation by \citet{2020AJ....160...92H}:
\begin{equation}
R_{\rm n} = \left[\dfrac{3 \kappa \mathscr{U}_{\rm gas} \mathfrak{m}_{\rm u}}{4 \pi \rho_{\rm n} \sqrt{\sum_{j=1}^{3} A_{j}^2}} \left(\dfrac{Q_{\rm gas} v_{\rm gas}}{g_{\rm NG}} \right) \right]^{1/3}
\label{eq:Rnuc_NG}.
\end{equation}
\noindent Here $\kappa$ is the collimation coefficient of activity, with a lower bound of 0 indicating an isotropic scenario and an upper bound of 1 corresponding to perfectly collimated outgassing, $\mathscr{U}_{\rm gas}$ is the relative molecular mass of the dominant outgassing substance, $\mathfrak{m}_{\rm u} = 1.66 \times 10^{-27}$ kg is the atomic mass constant, $\rho_{\rm n} = 0.5$ g cm$^{-3}$ is the assumed bulk density of the nucleus \citep[][and citations therein]{2019SSRv..215...29G}, $Q_{\rm gas}$ and $v_{\rm gas}$ are, respectively, the molecular production rate and the outflow speed of the dominant outgassing substance. The three terms inside the parentheses in Equation \eqref{eq:Rnuc_NG} are functions of heliocentric distance, while the others are constants or treated as such for simplicity.

We evaluated the nongravitational effect of 3I around the time of perihelion when its outgassing activity and the resulting nongravitational acceleration were most pronounced as a consequence of momentum conservation. The benefit of picking an epoch around perihelion is that the dominance of water production in the outgassing activity of 3I over other sublimating species, including CO$_{2}$, allowed us to safely neglect contributions from these various substances, which can no longer be omitted at greater heliocentric distances. At $r_{\rm H} = 1.401$ au, the interstellar object exhibited significant water production at a rate of $Q_{\rm gas} = \left(3.2 \pm 1.0 \right) \times 10^{29}$ s$^{-1}$, which subsequently declined during the outbound leg \citep{2025arXiv251222354C}.
\footnote{At the time of writing, \citet{2026arXiv260115443T} reported water production rates up to an order of magnitude lower than \citet{2025arXiv251222354C}, despite using the same dataset. While this difference deserves to be investigated, applying the correlation from \citet{2008LPICo1405.8046J} to the near-perihelion magnitudes of 3I reported by \citet{2025arXiv251025035Z} yielded water production rates of 3I in good agreement with \citet{2025arXiv251222354C} at similar heliocentric distances, albeit preperihelion, and so we do not adopt the results of \citet{2026arXiv260115443T}. }
Setting $\mathscr{U}_{\rm gas} = 18$ for water, $\kappa = 0.5$ to approximate hemispherical emission, and adopting the empirical law for outflow speed of gas from \citet{1993Icar..105..235C} yielding $v_{\rm gas} \approx 0.7$ km s$^{-1}$, we obtained the nucleus radius of 3I from Equation \eqref{eq:Rnuc_NG} to be $1.5 \pm 0.1 $ km, in excellent agreement with our results in Section \ref{ssec:nuc_ext} via the nucleus extraction technique, thereby solidifying our nucleus size estimate. 

Here the formal uncertainty was properly propagated from the covariance matrix of the RTN nongravitational parameters provided by Horizons and the reported error in the water production rate in \citet{2025arXiv251222354C}. 
However, this formal uncertainty  is a lower limit to the true uncertainty because it neglects systematic effects. The maximum nucleus size permitted by the detected nongravitational acceleration occurs under a fully collimated mass-loss scenario, i.e. $\kappa = 1$, which yields $R_{\rm n} \le 1.9 \pm 0.2$ km. Conversely, a much smaller nucleus size estimate will be inferred if $\kappa$ is much smaller, corresponding to less anisotropic mass loss. Furthermore, if a significant fraction of the measured water production is due to the sublimation of  icy grains in the coma rather than from the nucleus surface, our size estimates inferred from the nongravitational effect should be considered as upper limits to the actual nucleus radius.  We cannot reliably quantify the systematic uncertainties given the presently available data.

In Table \ref{tab:sz_nuc_comp}, we compare our nucleus size estimates of 3I with earlier results.  Our inferred nucleus radius is large compared to earlier results by  \citet{2025RNAAS...9..329E} and \citet{2025arXiv251218341F}, but small compared to that of \citet{2025arXiv250921408C}, despite all three groups applying the same methodology. We posit that earlier solutions for the nongravitational parameters were less reliable due to the shorter observed arc, where potential biases, such as tailward offsets in astrometric measurements, could be more easily absorbed into the orbital fits. Furthermore, the applied nongravitational force model might have been overly simplified and therefore failed to closely approximate earlier epochs when the outgassing of 3I was still weak and the nongravitational effect was far less significant. For instance, at greater heliocentric distances, the nongravitational acceleration could have been arisen from outgassing of multiple species rather than solely from the dominant volatile. The discrepancy with \citet{2025arXiv250921408C} stems from their reliance on an early upper limit for the nongravitational acceleration, which is inconsistent with the more recent Horizons solution by nearly two orders of magnitude, and to a lesser extent from their neglect of the collimation factor, $\kappa$.

\subsection{Statistics of Similar Interstellar Interlopers}
\label{ssec:stats_ISO}

We attempted to evaluate the statistics of interstellar interlopers similar in size to 3I  inside our solar system. Based on 1I/`Oumuamua alone, \citet{2017ApJ...850L..36J} and \citet{2018ApJ...855L..10D} derived a number density of $\sim\!0.1$-0.2 au$^{-3}$ for such objects in the solar system. The subsequent discovery of the larger 2I/Borisov was not a fluke but was in line with extrapolated statistics \citep{2020ApJ...888L..23J}. Recently, \citet{2025ApJ...989L..36S} reported a number density of $\sim\!3 \times 10^{-4} $ au$^{-3}$ for 3I-like objects with $H_{V} = 12.5$. However, because this absolute magnitude led to an overestimate of the nucleus size due to contamination from the coma, the real number density was likely underestimated. 

Using Figure 8 from \citet{2025ApJ...989L..36S}, we estimated the ratio of detectability volume to crossing time for the actual absolute magnitude of 3I's nucleus ($H_{{\rm n}, V} \approx 17.1$; see Section \ref{ssec:nuc_ext}) to be $\sim\!30$ au$^3$ yr$^{-1}$. The ratio of the ATLAS survey's detection rate ($\sim\!0.2$ yr$^{-1}$) to this volume-to-time ratio yields a local number density of $\sim\!7 \times 10^{-3}$ au$^{-3}$ for 3I-sized interstellar objects. Given the discovery of 3I at a heliocentric distance of $r_{\rm H} \approx 4.5$ au, this implies a total number of $\sim\!3$ similar interstellar objects within a sphere of this radius at the Sun at any instant on average. 

If extrapolated solely from the statistics based on 1I/`Oumuamua using a differential size distribution index of -3 for the population of interstellar objects, the expected total number of 3I-sized or larger interstellar objects inside the same volume of space around the Sun will be $\sim\!0.1$-0.3. The likelihood of detecting such an object is given by Poisson statistics to be $\sim\!0.1$-0.2, which drops to $\sim\!\left(1{\rm -}2\right)\%$ for a steeper size distribution of -4. However, these estimates should only be treated as lower limits because cometary activity undoubtedly enhances the detectability of 3I-like objects. We conclude that the discovery of 3I remains in qualitative agreement with the previous statistics of the interstellar object population. 

The arrival rate of 3I-like interstellar objects, estimated from the ratio of their total number within the 4.5 au radius sphere to the corresponding crossing timescale,
$\sim\!2 \times 10^{7}$ s, is $\sim\!4$ yr$^{-1}$ assuming a steady-state population. We found that the maximum crossing time for this distance would be $\sim\!1 \times 10^{8}$ s, corresponding to objects on barely hyperbolic trajectories, which implies a lower limit on the arrival rate of $\ga\!0.6$ yr$^{-1}$. A qualitatively similar result for the detection rate of 3I-sized objects, $\sim\!0.3$-0.6 yr$^{-1}$, can be reached by scaling from the relation presented in \citet{2022PSJ.....3...71H} using a differential size distribution index of -3, whereas a steeper index of -4 would yield a rate in the range of $\sim\!0.02$-0.04 yr$^{-1}$, smaller by an order of magnitude. 

Although all-sky surveys such as Pan-STARRS and Catalina have been operational for over a decade, only two other interstellar objects were previously discovered, both of which were visually fainter than 3I. Around its peak brightness, 3I was sufficiently bright that even visual comet hunters could have detected it had it passed through the inner solar system prior to the era of automated sky surveys. Here we simplistically estimate the likelihood that no 3I-like interstellar objects passed by between mid-1990s, the advent of modern sky surveys, and the discovery of 1I/`Oumuamua in 2017. Over this $\sim\!20$-year interval, the expected number of detections is $\sim\!4$, assuming a detection rate of $\sim\!0.2$ yr$^{-1}$. According to Poisson statistics, the probability of no detectable objects in this period is $\sim\!\exp \left(-4 \right) \approx 2\%$. Repeating this order-of-magnitude estimation by scaling from the relation in \citet{2022PSJ.....3...71H} with an index of -3 yields a probability of $\la\!3 \times 10^{-3}$, which is negligible. However, the probability of a null detection increases considerably to 0.5-0.6 if a steeper index of -4 is adopted. While earlier surveys were undeniably less sensitive if compared to contemporary ones, they were unlikely to be less efficient by many orders of magnitude. Therefore, if the size distribution of the interstellar object population is not steep, it appears highly probable that several 3I-like interstellar objects passed through the inner solar system undetected between the advent of automated sky surveys and the discovery of 1I/`Oumuamua. Alternatively, the lack of prior detections may imply a steep size distribution for the population, potentially in agreement with the inference in \citet{2017ApJ...850L..36J}.

\subsection{Dust Activity \& Scattering Phase Function}
\label{ssec:act}

\begin{deluxetable*}{ccccccc}
\tablecaption{
Best-Fit Postperihelion Lightcurve Models of 3I/ATLAS
\label{tab:act}}
\tablewidth{0pt}
\tablehead{
\colhead{Aperture Radius} &
\colhead{Absolute Magnitude} &
\colhead{Activity Index} &
\colhead{Linear Phase Slope} &
\colhead{Opposition Effect} &
\colhead{Opposition Effect} &
\colhead{Normalised}
\\
(km) &
$H_{{\rm c}, V}$
 &
$n$
 &
$\beta_{\alpha}$ (mag deg$^{-1}$)
 &
Magnitude $m_{\rm OE}$
 &
Width $w_{\rm OE}$ (\degr) 
&
Residual RMS\tablenotemark{$\dagger$}
}
\startdata
200 & $11.5 \pm 0.7 $ & $4.8 \pm 0.6$ & $0.03 \pm 0.01$ & $0.15 \pm 0.02$ & $2 \pm 2$ & 0.551 \\
400 & $11.0 \pm 0.9$ & $4.5 \pm 0.7$ & $0.03 \pm 0.01$ & $0.21 \pm 0.03$ & $3 \pm 2$ & 0.137 \\
600 & $10.8 \pm 1.0$ & $4.3 \pm 0.7$ & $0.02 \pm 0.01$ & $0.22 \pm 0.03$ & $4 \pm 2$ & 0.087 \\
800 & $10.5 \pm 1.0$ & $4.3 \pm 0.7$ & $0.02 \pm 0.01$ & $0.23 \pm 0.03$ & $4 \pm 2$ & 0.060 \\
1000 & $10.2 \pm 0.9$ & $4.3 \pm 0.7$ & $0.02 \pm 0.01$ & $0.23 \pm 0.03$ & $3 \pm 2$ & 0.048 \\
\hline
\multicolumn{1}{l}{Mean} & -- &
$4.5 \pm 0.3~\left(0.2 \right)$ &
$0.026 \pm 0.006~\left(0.005 \right)$ & 
$0.19 \pm 0.01~\left(0.04 \right)$ &
$3 \pm 1~\left(1 \right)$ &
--
\enddata
\tablenotetext{\dagger}{Root mean square of the observed-minus-calculated residual sigma.
}
\tablecomments{The presented mean values are weighted means computed from measurements across all apertures, followed by their associated standard errors and standard deviations listed in parentheses.}

\end{deluxetable*}


\begin{figure*}
\centering
\begin{subfigure}[b]{.49\textwidth}
\centering
\includegraphics[width=\textwidth]{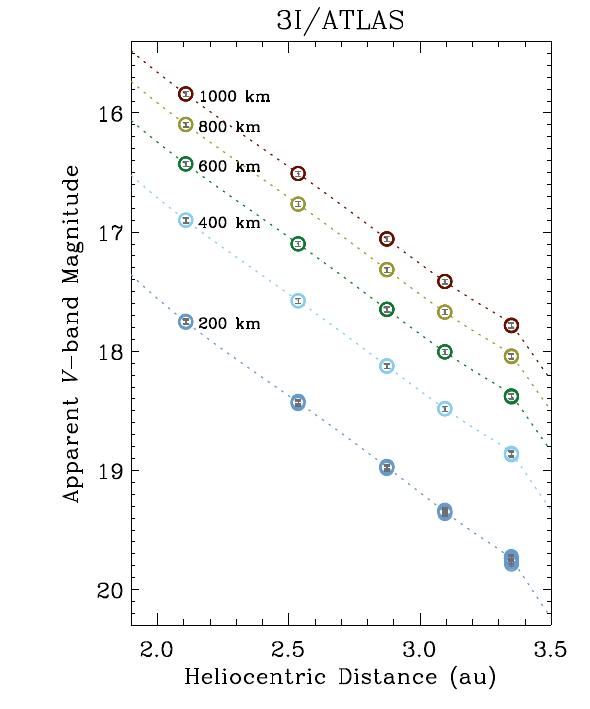}
\caption{
\label{fig:lc}
}
\end{subfigure}
\begin{subfigure}[b]{.49\textwidth}
\centering
\includegraphics[width=\textwidth]{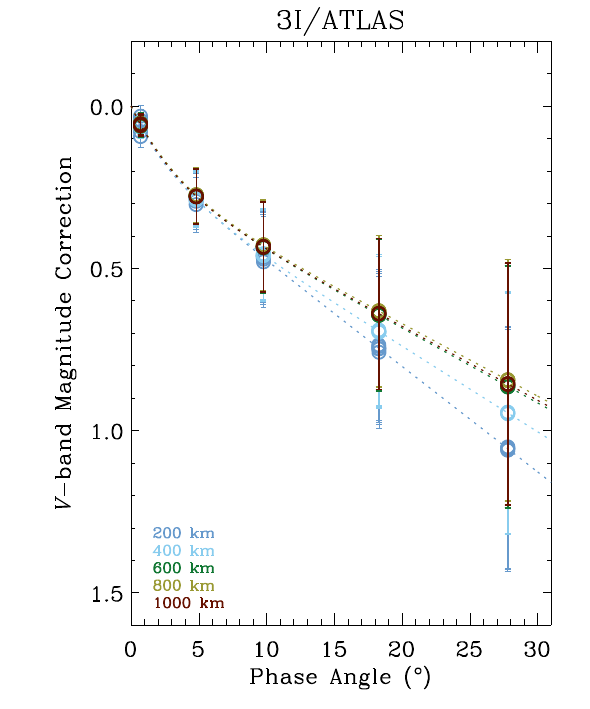}
\caption{
\label{fig:phi}
}
\end{subfigure}
\caption{
Left: Apparent $V$-band magnitude of postperihelion 3I/ATLAS as a function of heliocentric distance across different apertures of fixed linear radii (colour coded and labelled).
Right: $V$-band magnitude correction versus phase angle, highlighting the opposition effect. 
Best-fit models incorporating the linear-exponential phase function are drawn as dotted curves.
\label{fig:coma}
}
\end{figure*}

Aside from characterising the nucleus of 3I, we analysed its postperihelion activity and the scattering properties of its dust grains using HST observations covering a month-long baseline and a wide range of phase angles (see Table \ref{tab:vgeo}). Similar to the photometric procedure in Section \ref{ssec:cc}, we performed aperture photometry at the photocenter of the interstellar interloper but employed multiple apertures with constant linear radii at the projected HST-centric distance of 3I so as to sample the same cross-section regardless of varying distances. We used fixed linear radii from 200 km to 1000 km in 200 km increments. The sky background was measured from a concentric annular region with radii extending from $\sim\!32\arcsec$ to 79\arcsec~(800-2000 pixels) to minimise contamination from the dust environment of 3I. Using annuli further from the photocenter did not materially alter the results but instead increased uncertainty because reduced area within the FOV of the WFC3 camera fell inside the annuli.

We modelled the apparent magnitude of 3I as a function of heliocentric distance, HST-centric distance, and phase angle, following the formalism of Equation \eqref{eq:Hnuc}:
\begin{equation}
m_{{\rm c}, V} = H_{{\rm c}, V} + 2.5 n \log r_{\rm H} + 5 \log \Delta + \beta_{\alpha} \alpha
\label{eq:Hcom},
\end{equation}
\noindent in which $H_{{\rm c},V}$ is technically the heliocentric magnitude of 3I at $r_{\rm H} = 1$ au but is still termed absolute magnitude here, and $n$ is the activity index parametrising the heliocentric dependency of activity, with $n = 2$ corresponding to a constant effective cross-section. These two parameters, along with the linear phase slope $\beta_{\alpha}$, which was held fixed for the nucleus in Section \ref{sec:an}, were treated as free parameters of the model. Utilising the Levenberg-Marquardt approach in {\tt MPFIT}, we obtained best-fit parameters and their uncertainties derived from the $3 \times 3$ covariance matrix, properly propagated from the individual measurement errors. The best fits yielded weighted mean values $\bar{n} = 4.7 \pm 0.2$ and $\bar{\beta}_{\alpha} = 0.035 \pm 0.003$ mag deg$^{-1}$ for the activity index and linear phase slope, where the uncertainties represent standard errors, as the values remained statistically consistent across aperture sizes. However, the best fits all yielded reduced $\chi^2$ values $\chi_{\nu}^{2} > 2$, implying residuals significantly greater than the measurement uncertainties. By examining the resulting magnitude correction with respect to the absolute magnitude as a function of phase angle (hereafter phase function for conciseness, and denoted $\phi$), we clearly detected what appears to be an opposition surge in 3I, deviating from the trend predicted by the purely linear phase function.

Therefore, we refitted the apparent magnitude of 3I by replacing the fourth term in Equation \eqref{eq:Hcom} with the modified linear-exponential phase function from \citet{Rosenbush2002}:
\begin{equation}
\phi\left(\alpha\right) = \beta_{\alpha} \alpha + m_{\rm OE} \left[1 - \exp\left(- \dfrac{\alpha}{w_{\rm OE}} \right) \right]
\label{eq:phi_LEPF},
\end{equation}
\noindent where $\beta_{\alpha}$ is the slope of the linear portion of the phase function, $m_{\rm OE}$ is the opposition surge in magnitude, and and $w_{\rm OE}$ characterises the $e$-folding width of the opposition effect. The inclusion of these two additional free parameters considerably improved the reduced $\chi^2$ values by at least an order of magnitude, and by up to a factor of $\sim\!10^{3}$ for several apertures. We compare our measurements against the best-fit models in Figure \ref{fig:coma}) and tabulate the best-fit parameters in Table \ref{tab:act}), where mean values except for the absolute magnitude along with the respective standard errors and standard deviations are also appended, in that the best-fit absolute magnitude of 3I brightens with growing aperture size as expected as a consequence of more effective cross-section of dust grains in the coma being measured, while the other best-fit parameters remain consistent across different apertures.

The use of an alternative phase function model yielded a mean activity index of $\bar{n} = 4.5 \pm 0.03$, which is not statistically distinguishable from the one with the linear model, implying that the result of the activity index is robust and relatively independent of the chosen phase correction. While preperihelion continuum photometry from $r_{\rm H} = 4.6$ au to 1.8 au revealed a steady brightening of 3I well described by an activity index  $n = 3.8 \pm 0.3$ \citep{2025ApJ...994L...3J}, our postperihelion result implies that, over similar ranges of heliocentric distance, 3I faded more rapidly on the outbound leg than it brightened on the inbound leg.\footnote{Overlapping satellite observations showed a  more rapid brightening towards perihelion \citep{2025arXiv251025035Z}, but sampled a mixture of gas and dust continuum emission that cannot be meaningfully interpreted; we ignore them here.} This activity asymmetry is further corroborated by the fact that the coma of 3I exhibited a significantly steeper surface brightness gradient preperihelion \citep[$\gamma > 1$;][]{2025ApJ...990L...2J, 2025ApJ...994L...3J} than the shallower postperihelion profile with $\gamma < 1$ found in Section \ref{ssec:nuc_ext} (see Figure \ref{fig:coma_pars}). The heliocentric index is influenced by many unmeasured parameters of the nucleus, among them the distribution of surface volatiles and the direction of the nucleus spin vector.  It is thus impossible to ascribe a unique interpretation to the measured change. We note, however, that measurements of solar system comets show that the heliocentric index can decrease or increase after perihelion with equal probability, and that the steeper post-perihelion index of 3I cannot be regarded as unusual \citep{2025A&A...697A.210L}.  


Finally, we briefly comment on the scattering phase function of 3I. Thanks to the smooth decline in activity and a range of phase angles extending from almost $\sim\!30\degr$ all the way to near-opposition ($\alpha \approx 0\degr$), we were able not only to robustly constrain the linear part of the phase function, with a slope of $\bar{\beta}_{\alpha} = 0.026 \pm 0.006$ mag deg$^{-1}$, but also its opposition surge, neither of which had been previously measured for 3I. The linear portion of the scattering phase function of 3I appears indistinguishable from those of solar system comets. In contrast, to date there remains a persistent dearth of intensity measurements for cometary dust near zero phase angle \citep{Rosenbush2002}. Among the few comets that have been studied in this regime, none have exhibited a distinct opposition effect \citet{1987A&A...187..585M}. This phenomenon, however, is widely observed among asteroids, planetary rings, and miscellaneous other airless bodies in the solar system \citep{Rosenbush2002}. While we acknowledge that the light scattering mechanisms of cometary dust differ from those of airless surfaces, the opposition effect of 3I qualitatively resembles some of the latter. Although \citet{1987A&A...187..585M} found no opposition effect in their selected sample of solar system comets, the brightness enhancement of 3I we observed is broadly in line with their reported upper limit for the opposition surge of $m_{\rm OE} \la 0.2$ mag. 

While we cannot rule out the possibility from the available data that the dust grains of 3I are inherently strong backscatterers, we suspect that the detectability of this effect might have been significantly enhanced by concurrent contributions from the orbital plane crossing and the projection of the dust tail. Both factors would increase the total effective cross-section of dust within the photometric aperture, for which we did not correct in Equation \eqref{eq:Hcom}. We leave the disentanglement of these competing factors to future dust-modelling work. Taken together with the peculiar polarimetric properties reported by \citet{2025ApJ...992L..29G,2026arXiv260108591C}, our results  suggest that while there are commonalities between dust grains of 3I and solar system comets, unusual physical properties exist. 

We did not attempt to constrain the nucleus scattering properties of 3I because of the unconstrained rotation lightcurve modulation from our sparse sampling of the HST observations. Although the near opposition observation from Visit 6 may be influenced by an opposition effect of the nucleus, we posit that the adopted linear phase function typical for cometary nuclei in the solar system for the nucleus of 3I possibly remains valid, given our practice with fitting for the scattering properties of its cometary dust.


\section{Summary}
\label{sec_sum}

This paper presents a photometric analysis of our Hubble Space Telescope observations of interstellar object 3I/ATLAS on the outbound leg of its heliocentric trajectory. Our key findings are as follows:
\begin{enumerate}
    \item Using the nucleus extraction technique, we successfully revealed the nucleus of 3I and measured its optical cross-section (the product of geometric albedo and physical cross-section) to be $C_{\rm n} = 0.22 \pm 0.07$ km$^{2}$. Adopting a nominal $V$-band geometric albedo of 0.04, typical for cometary nuclei in the solar system, we estimate an effective nucleus radius of $R_{\rm n} = 1.3 \pm 0.2$ km. The result is  consistent with our independent estimate derived from the reported nongravitational acceleration and production rates of 3I.
    \item The nucleus lightcurve exhibits evidence of temporal variations, likely attributable to either the rotation modulation of an aspherical shape or fluctuations in near-nucleus activity, or both. If due to rotation, the observed brightness range implies a nucleus axis ratio of $2:1$ in the plane of the sky. Unfortunately, none of our HST visits monitored a more complete phase of the lightcurve, making this dataset alone likely not useful for constructing a full phase lightcurve. Using the lightcurve trends we constrain the rotation period to be $\ga\!1$ hr.
    \item Scattered light by dust grains in the coma displayed a statistically significant opposition surge of $\sim\!0.2$ mag, characterised by an $e$-folding width of $3\degr \pm 1\degr$. This brightening possibly contain coincident contributions from the orbital plane crossing and tail projection. While opposition surges have not been widely reported for solar system cometary dust, the observed brightening is broadly consistent with existing upper limits for solar system comets. The linear portion of the phase function, with a slope of $\bar{\beta}_{\alpha} = 0.026 \pm 0.006$ mag degree$^{-1}$, is indistinguishable from those of solar system comets.

    \item Our fixed linear radius aperture photometry indicates that the postperihelion activity of 3I declined smoothly and  more rapidly than its preperihelion brightening trend, consistent with our finding that the postperihelion coma surface brightness profile was shallower than its preperihelion counterpart. We measured a postperihelion activity index of $n = 4.5 \pm 0.3$. Such asymmetries in 
    brightening and fading rates are common among solar system comets.


    \item Based on the discovery statistics of interstellar objects, we estimate a minimum of three 3I-sized (nucleus radius $\sim\!1.3$ km) interstellar object within $r_{\rm H} = 4.5$ au at any given instant, which is likely a conservative lower bound as inactive interstellar objects of this size would be significantly more difficult to detect than an active object like 3I. 
    It is probable that  comparably bright interstellar objects have passed through the inner solar system during the era of wide-field CCD surveys; we surmise that such objects have  been missed. Alternatively, the lack of prior detections may imply a steep size distribution for the interstellar object population.
    
\end{enumerate}

\begin{acknowledgments}

We thank 
Alison Vick for her great efforts in scheduling the HST observations and ensuring the successful execution of the HST program.
We also thank Darryl Seligman and Maria Womack for their prompt and insightful reports, which was helpful in improving our manuscript.
This research is based on observations made with the NASA/ESA Hubble Space Telescope obtained from the Space Telescope Science Institute, which is operated by the Association of Universities for Research in Astronomy, Inc., under NASA contract NAS 5–26555. No NASA funding was used for this work. These observations are associated with program GO/DD 18152.

\end{acknowledgments}




\facilities{HST (UVIS)}

\software{
{\tt IRAF} \citep{1986SPIE..627..733T},
{\tt L.A.Cosmic} \citep{2001PASP..113.1420V}, 
{\tt MPFIT} \citep{2009ASPC..411..251M},
{\tt psffit} (\href{https://www2.boulder.swri.edu/~buie/idl/}{Buie's IDL library})}

\bibliography{3I_HST}{}
\bibliographystyle{aasjournalv7}

\end{CJK}
\end{CJK}

\end{document}